\documentclass[preprint,aps,prb,amssymb,floatfix,showpacs,showkeys,
]{revtex4-2}
\usepackage{amsmath}
\usepackage{graphicx}
\usepackage[dvips]{epsfig}
\usepackage{array}
\usepackage{romannum}
\AtBeginDocument{\pagenumbering{arabic}}

\usepackage{hyperref}

\usepackage{color}
\usepackage{amsthm}

\newcommand{\Ron}{\Romannum}

\allowdisplaybreaks

\newtheoremstyle{query}%
{}{}
{\color{red}}
{}
{\sffamily\bfseries}{:}{12pt}
{}
\theoremstyle{query}
\newtheorem{aq}{Author Query/Comment}

\newcommand{\AQ}[1]{{\textcolor{red}{#1}}}

\newcommand{\baq}{\begin{aq}}
\newcommand{\eaq}{\end{aq}}

\begin{document}

\title{Electronic properties of bilayer graphene with magnetic quantum structures studied using the Dirac equation }

\author{Daehan Park}
	\affiliation{Department of Physics, Soongsil University, Seoul 06978, Korea}
\author{Heesang Kim}
 	\affiliation{Department of Physics, Soongsil University, Seoul 06978, Korea}
\author{Nammee Kim}
	\email{nammee@ssu.ac.kr}
	\affiliation{Department of Physics, Soongsil University, Seoul 06978, Korea}

\date{\today}

\begin{abstract}
The electronic properties of bilayer graphene with a magnetic quantum dot and a magnetic quantum ring are investigated. The eigenenergies and wavefunctions of quasiparticle states are calculated analytically by solving  decoupled fourth-order differential equations.
For  the magnetic quantum dot, in the case of a negative inner magnetic field, two peculiar characteristics  of the eigenenergy evolution are found: (i) the  energy eigenstates change in a stepwise manner owing to energy anticrossing and (ii) the quantum states approach  zero energy.
For the magnetic quantum ring, there is an angular momentum transition of eigenenergy as the inner radius of the ring varies,  and the Aharonov--Bohm effect is observed in the  eigenenergy spectra for both positive and negative magnetic fields inside  the inner radius.
\end{abstract}


\maketitle


\section{Introduction}

Since the first demonstration of the fabrication of monolayer graphene~\cite{Science_5696_666_Novoselov}, graphene has attracted considerable
attention owing to its exotic properties, such as its long mean free path, high carrier mobility, and an anomalous  quantum Hall effect~\cite{Science_5696_666_Novoselov, Nature_438_197_Novoselov2005, APL_99_232104_Zomer,PhysRevLett_95_146801_Gusynin, Nature_438_201_Zhang}.
These properties of graphene   originate mainly from the existence of two independent Dirac points, which are conventionally called the $K$ and $K'$ valleys. In
the vicinity of these valleys, graphene has a unique band structure, namely, a gapless and linear energy dispersion. As a result, electrons in graphene behave as if they were
massless relativistic  particles, governed by the Dirac equation.

However, these relativistic characteristics are a considerable hindrance to the realization of electronic devices based on graphene.
Owing to the Klein effect~\cite{Klein1929}, near-perfect transmission of Dirac electrons through an electrostatic barrier at normal incidence  is possible.
That is, unlike electrons in a two-dimensional electron gas (2DEG), Dirac electrons cannot be confined by an electrostatic potential. However, De Martino \emph{et al.}~\cite{PhysRevLett_98_066802_Martino}
showed that this problem could be overcome in graphene by the application of a magnetic field. In contrast to an electrostatic potential, by deflecting the trajectory of electrons through the Lorentz force,  a magnetic field allows trapping of electrons
 within a restricted region. Various magnetic quantum structures such as magnetic quantum dots, magnetic step barriers, and
magnetic quantum rings with impurities have been investigated
~\cite{PhysRevLett_98_066802_Martino, SolidStateComm_144_12_Martino, PhysLettA_373_4082_WANG, PhysRevB_79_155451_Masir, JPSJ_83_034007_Lee}.

Graphene multilayers  have also been widely investigated. In multilayer graphene, the layers are weakly coupled by van
der Waals interaction, and its properties are very different from those of monolayer graphene. For instance,  pristine bilayer
graphene (BLG) exhibits gapless and parabolic energy dispersion near the $K$ and $K'$ valleys~\cite{Rep_Prog_Phys_76_056503_McCann}.
Also, in a uniform magnetic field $B$, the Landau levels (LLs) of bilayer graphene in the lowest band have a linear dependence on $B$, in contrast to the
 $\sqrt{B}$ dependence of LLs in monolayer graphene~\cite{PhysRevB_78_033403_Nakamura}.
%
%
%
%
%

In recent years, semiconductor-based quantum dots and quantum rings have become important subjects of study in mesoscopic
physics~\cite{Nature_413_822_Fuhrer_2001}. These quantum structures have been investigated both experimentally and theoretically in
BLG. There have been theoretical studies of quantum dots and quantum rings in AA-stacked~\cite{Mat_Research_express_Zahidi_2017, Mat_Research_express_Belouad_2016,PhysRevB_102_075429},
 AB-stacked~\cite{NanoLett_7_946_Pereira2007, PhysRevB_79_195403_Pereira,NanoLett_9_4088_Zarenia, PhysRevB_81_045431_Zarenia, Carbon_78_392_Costa, PhysRevB_93_085401}, and twisted \cite{PhysicaE_112_36_Tiutiunnyk} BLG.
The properties of  two or more coupled quantum dots in BLG have also been studied experimentally and theoretically~\cite{acs_nanolett_20_2005, acs_nanolett_18_5042, PhysRevB_96_035434}.
Recently, the  electronic density within BLG
 quantum dots has been visualized using the tip of a scanning tunneling microscope (STM)~\cite{acs_nanolett_20_8682_Zhehao,acs_nanolett_18_5104_Velasco}.
The valley properties of an electrostatically confined BLG quantum dot have been investigated experimentally
~\cite{PhysRevLett_123_026803, PhysRevX_8_031023}, and a valley filter device based on these properties has been proposed~\cite{PhysRevB_99_125422}.
%
%
%
%

In this paper, based on a single-particle approach, we investigate the electronic properties of BLG with two kinds of magnetic
structures: a magnetic quantum dot (MQD) and a magnetic quantum ring (MQR).
In most  previous work on  these quantum structures in a 2DEG and in monolayer graphene, there was no magnetic field inside the MQD, and the spatial area of the MQR was kept constant.
Here, however,  we consider the behavior of the eigenenergies of BLG as the magnetic
field inside an MQD varies. We show that there are dramatic differences depending on the direction of the magnetic field  inside the dot.
The results for the eigenenergy spectra of BLG for an MQR with fixed spatial area  reveal nothing unusual compared with the cases of a 2DEG and monolayer graphene. Therefore, we vary the inner radius of the ring and introduce antiparallel magnetic fields  inside the inner circle and outside the outer circle of the ring to look for interesting phenomena that might arise.

The remainder of the paper is organized as follows. In Sec.~\ref{secII}, we introduce a mathematical model of our magnetic quantum structures and present the mathematical solution procedure for the Hamiltonian of the
BLG, including complicated boundary conditions for the fourth-order differential equations satisfied by the Dirac wavefunctions. In Sec.~\ref{secIIIA}, we present our  results for BLG with an MQD and discuss its energy anticrossing by using  effective potentials and the  probability density at subatoms.
We  discuss the eigenenergies and the Aharonov--Bohm effect in BLG with an MQR in
Sec.~\ref{secIIIB}. Finally, we summarize our results in Sec.~\ref{secIV}.


\section{MODEL AND FORMALISM}\label{secII}

\begin{figure}
	\includegraphics[width = 8.5cm]{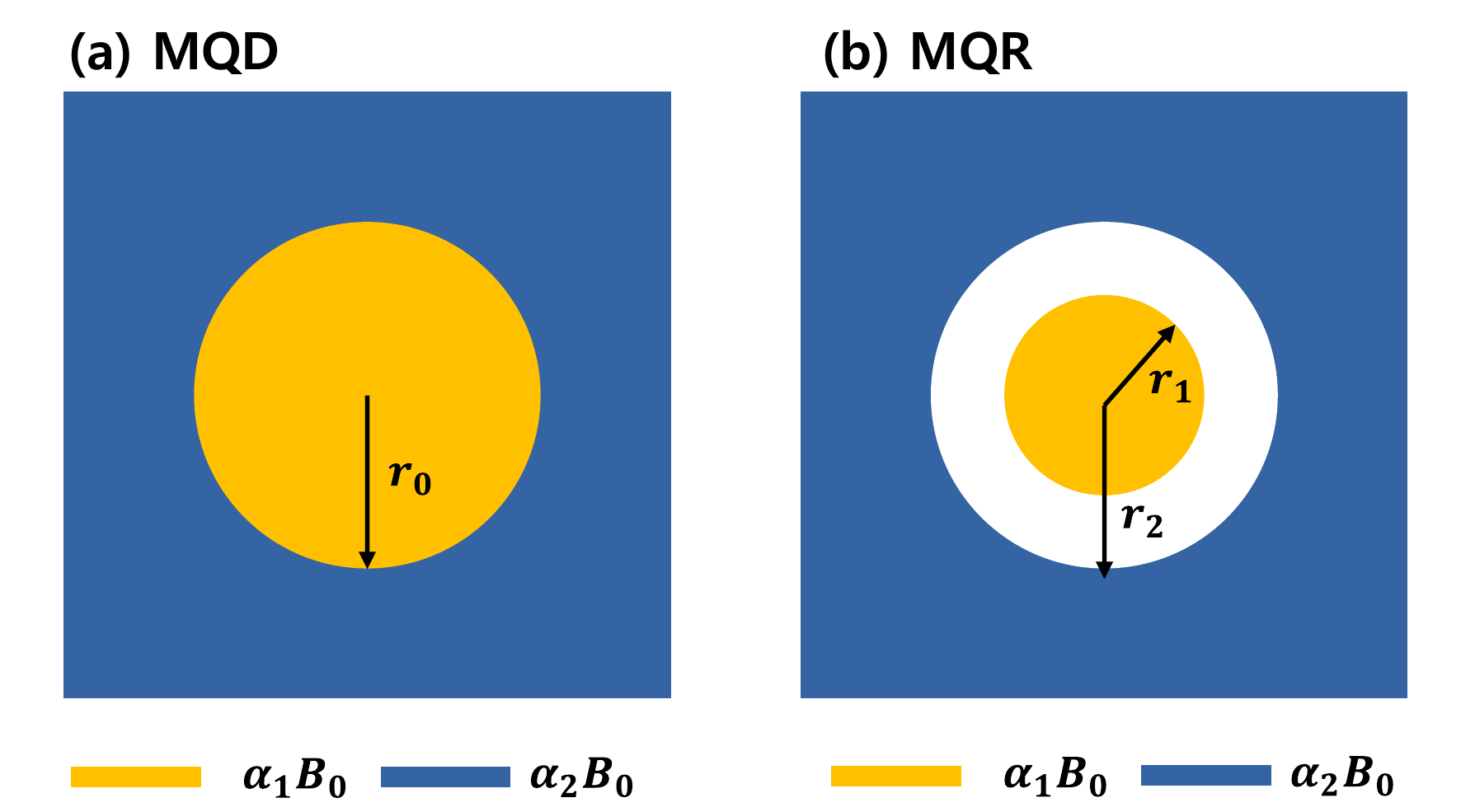}
	\caption{Schematic geometry of (a) a magnetic quantum dot (MQD) and (b) a magnetic quantum ring (MQR).\label{fig1}}
\end{figure}

We calculate the energy spectra of BLG with an MQD or an MQR, formed by a spatially nonuniform distribution of magnetic fields, by solving the Dirac equation.
The  geometries of the MQD and MQR are shown schematically in Fig.~\ref{fig1}. In Fig.~\ref{fig1}(a), the MQD is modeled as
$\vec{B}=\alpha_{1}B_{0}\hat{e}_{z}$ for $r<r_0$ and $\vec{B}=\alpha_{2}B_{0}\hat{e}_{z}$ for $r > r_0$, where $\alpha=B/B_{0}$ is the
ratio of the applied magnetic field $B$ to the standard magnetic field $B_0$. The MQDs studied previously correspond to  $\alpha_1=0$.
In Fig.~\ref{fig1}(b), the MQR is modeled as $\vec{B}=\alpha_{1}B_{0}\hat{e}_{z}$ for $r<r_{1}$, $\vec{B}=0$ for $r_{1} < r < r_{2}$, and
$\vec{B}=\alpha_{2}B_{0}\hat{e}_{z}$ for $r>r_{2}$.
The model of  the MQD can be obtained from that for the MQR in  the limit as $r_1\rightarrow r_2$. Therefore, we shall  describe the calculations for the MQR only.
Because the MQR has rotational symmetry, it is convenient to use  plane polar coordinates in our system.
We can express the vector potential $\vec{A}(\mathbf{r})$ of the magnetic field $\vec{B}$ in the symmetric gauge as
\begin{equation}
\vec{A}(\mathbf{r}) =
    \begin{cases}
		\dfrac{1}{2}\alpha_{1} B_{0} r \hat{\theta}&         \text{for $r<r_{1}$} ,\\[9pt]
		\dfrac{1}{2r}\alpha_{1} B_{0} r_{1}^2 \hat{\theta}&  \text{for $r_{1}<r<r_{2}$},\\[9pt]
		\dfrac{1}{2}\alpha_{2} B_{0} r \hat{\theta} - \dfrac{B_0}{2r}(\alpha_{2}r_{2}^2 - \alpha_{1}r_{1}^2)\hat{\theta} &\text{for $r>r_{2}$}.
    \end{cases}
\end{equation}

The  Hamiltonian for  BLG with Bernal (AA) stacking near the $K$ valley is given by~\cite{Rep_Prog_Phys_76_056503_McCann}
\begin{equation}\label{eq2}
	H =
	\begin{pmatrix}
		0            & \pi & t_\perp   & 0             \\
		\pi^{\dagger}&   0 & 0         & 0             \\
		t_\perp	     &   0 & 0         & \pi^{\dagger} \\
		0            &   0 & \pi       & 0
	\end{pmatrix},
\end{equation}
where  $\pi = \Pi_{x} + i\Pi_{y}$, $\pi^\dagger  = \Pi_{x} - i\Pi_{y}$, $\Pi_{i}= v_{F}[p_{i} + (e/c)A_{i}]$,
the Fermi velocity $v_{F} \approx c/300$, $t_\perp=400$~meV is an interlayer hopping parameter, and $c$ is the speed of light. The wavefunctions of
the Hamiltonian in Eq.~\eqref{eq2} can be expressed as a four-spinor wavefunction $\vec{\psi} = (\psi_{A}, \psi_{B}, \psi_{B'}, \psi_{A'})^{T}$.
In our system, the
vector potentials have spatial circular symmetry, and thus this four-spinor wavefunction can be written as~\cite{NanoLett_7_946_Pereira2007}
\begin{equation}
	\vec{\psi} = e^{i m \theta}\!\left(\phi_{A}(r), ie^{-i\theta}\phi_{B}(r), \phi_{B'}(r), ie^{i\theta} \phi_{A'}(r)\right)^{T},
\end{equation}
where $m$ is an angular momentum quantum number. To simplify our problem, we  nondimensionalize  quantities by taking the unit of length  to be
the magnetic length  $l_{B} = \sqrt{2\hbar/eB_{0}}$ and the unit of energy to be $\epsilon_{0}=\sqrt{2}\,\hbar v_{F}/l_{B}$.
For a standard magnetic field $B_{0}=10$~T, the values of $l_{B}$ and $\epsilon_{0}$ are $11.4$~nm and $81.05$~meV, respectively.

In dimensionless units, by solving the Dirac equation, $(H-E)\vec{\psi} = 0$, we  obtain the following four coupled first-order differential equations:
\begin{align}
	\left[\frac{d}{dr} - \frac{m-1}{r} - v(r)\right]\!\phi_{B}   &=   \sqrt{2}\,(\epsilon \phi_{A} - t\phi_{B'}), \label{eq4}\\[9pt]
	\left[\frac{d}{dr} + \frac{m  }{r} + v(r)\right]\!\phi_{A}   &= - \sqrt{2}\,\epsilon \phi_{B},\label{eq5}                \\[9pt]
	\left[\frac{d}{dr} + \frac{m+1}{r} + v(r)\right]\!\phi_{A'}  &=   \sqrt{2}\,[\epsilon \phi_{B'}(r) - t\phi_{A}],\label{eq6}\\[9pt]
	\left[\frac{d}{dr} - \frac{m  }{r} - v(r)\right]\!\phi_{B'}  &= - \sqrt{2}\,\epsilon \phi_{A'},\label{eq7}
\end{align}
where $t=t_\perp/\epsilon_0=4.96$. The expressions for $v(r)$ in each region are as follows:
\begin{equation}
	v(r) =
 		\begin{cases}
		 		\alpha_1 r                                           & \text{for $r<r_{1}$},       \\[9pt]
		 		\dfrac{\alpha_1 r_{1}^2}{r}                          & \text{for $r_{1}<r<r_{2}$}, \\[9pt]
		 		\alpha_2r - \dfrac{\alpha_2r_2^2 - \alpha_1r_1^2}{r} & \text{for $r_{2}<r$}.
    	\end{cases}
\end{equation}
Therefore, we need to solve  four coupled first-order differential equations. Fortunately, however, Eqs~\eqref{eq4}--\eqref{eq7} can be
decoupled to give a fourth-order differential equation for $\phi_{A}(r)$ in each region:
\begin{align}
\intertext{in region \Ron{1} ($r < r_1$),}
	\left[\frac{d^2}{dr^2} + \frac{1}{r}\frac{d}{dr} - \frac{m^2}{r^2} - 2\alpha_1 m - \alpha_{1}^2 r^2 + 2\epsilon^2 \right]^2\! \phi^{\Ron{1}}_{A}
	&= 4(\alpha^2+t^2\epsilon^2)\phi^{\Ron{1}}_{A},\label{eq9}\\
\intertext{in region \Ron{2} ($r_1<r<r_2$),}
	\left[\frac{d^2}{dr^2} + \frac{1}{r}\frac{d}{dr} - \frac{m_2^2}{r^2} + 2\epsilon^2 \right]^2\!\phi^{\Ron{2}}_{A}&= 4t^2\epsilon^2\phi^{\Ron{2}}_{A},\label{eq10}\\
\intertext{in region \Ron{3} ($r_2<r$),}
	\left[\frac{d^2}{dr^2} + \frac{1}{r}\frac{d}{dr} - \frac{m_3^2}{r^2} - 2\alpha_2 m_3 - \alpha_{2}^2 r^2 + 2\epsilon^2
	\right]^2\!\phi^{\Ron{3}}_{A}& = 4(\alpha^2+t^2\epsilon^2)\phi^{\Ron{3}}_{A}, \label{eq11}
\end{align}
where $m_2$ and
$m_3$ are $m + \alpha_{1} s_{1}$ and $ m - (\alpha_{2} s_{2} - \alpha_{1} s_{1})$, respectively. Here, $\alpha_1 s_{1} = \pi\alpha_1  B_{0}
r_{1}^2/\phi_0$ and $\alpha_2 s_2 = \pi \alpha_2 B_{0} r_2^2/\phi_0$ are the numbers of
magnetic flux quanta inside $r< r_1$ and $r< r_2$, respectively, where $\phi_0 = h/e$ is a flux quantum and  $r_{i} = \sqrt{s_{i}}$ in dimensionless units.

The solutions of Eqs.~\eqref{eq9}--\eqref{eq11} for $\phi_{A}$ are
\begin{widetext}
	\begin{equation}
	\phi_{A}(r) =
   	 	\begin{cases}
			r^{|m|} e^{ - \alpha_1 r^2/2} [
			  A M(\delta_{+}^{1} ,|m| + 1, \alpha_{1} r^2) + B M(\delta_{-}^{1} ,|m| + 1, \alpha_{1} r^2) ]
			& \text {for $r < r_1$}, \\[6pt]
			C_{1} J_{|m_{2}|} (\sqrt{2}\gamma_{+}r) + D_{1} N_{|m_{2}|} (\sqrt{2}\gamma_{+}r)
			+
			C_{2} J_{|m_{2}|} (\sqrt{2}\gamma_{-}r) + D_{2} N_{|m_{2}|} (\sqrt{2}\gamma_{-}r) & \text{for $r_1<r<r_2$}, \\[6pt]
			r^{|m_{3}|} e^{ -\alpha_2 r^2/2} [
			  E U(\delta_{+}^2, |m_{3}| + 1, \alpha_{2} r^2) +
			  F U(\delta_{-}^2, |m_{3}| + 1, \alpha_{2} r^2)]& \text{for $r_2 < r$},
    \end{cases}
	\end{equation}
\end{widetext}
where
 \[
 \delta_{\pm}^{i} = - \frac{\beta_{i,\pm}^2}{2 \alpha_{i}} + \frac{1}{2}\!\left (|m| + \frac{\alpha_i}{|\alpha_i|}m + 1\right), \quad
 \beta_{i,\pm} = \sqrt{ \epsilon^2 \pm \sqrt{\alpha_{i}^2 + (t\epsilon)^2}}\,, \quad \gamma_{\pm} =  \sqrt{\epsilon^2 \pm t\epsilon},
 \]
 $M$ and $U$ are  confluent
hypergeometric functions, and $J_m$ and $N_m$ are Bessel functions of the first and second kinds, respectively. The wavefunctions
for the other subatoms, namely, $\phi_B$, $\phi_{A'}$, and $\phi_{B'}$, can also be derived from  Eqs.~\eqref{eq4}--\eqref{eq7}. Unlike the Schr\"odinger equation and
the Dirac equation for monolayer graphene, the  equations for $\phi_A$ for BLG are fourth-order differential equations. Therefore, we need to derive appropriate boundary
conditions for $\phi_A$. These boundary conditions can be obtained from the continuity of the wavefunctions for all subatoms, $\phi_A$,
$\phi_B$, $\phi_A'$, and $\phi_B'$, at the boundaries as  follows:
\begin{align}
\intertext{at $r = r_1$, which is the boundary between regions \Ron{1} and \Ron{2},}
	\phi_{A}^{\Ron{1}}(r_1) - \phi_{A}^{\Ron{2}}(r_1) &= 0  \\[6pt]
	\left.\frac{d}{dr}\phi_{A}^{\Ron{1}}\right\rvert_{r_1} - \left.\frac{d}{dr}\phi_{A}^{\Ron{2}}\right\rvert_{r_1} &= 0,  \\[6pt]
	\left.\frac{d^2}{dr^2}\phi_{A}^{\Ron{1}}\right\rvert_{r_1} + 2\alpha_1\phi_{A}^{\Ron{1}}(r_1) - \left.\frac{d^2}{dr^2}\phi_{A}^{\Ron{2}}\right\rvert_{r_1}  &= 0, \\[6pt]
	\left.\frac{d^3}{dr^3}\phi_{A}^{\Ron{1}}\right\rvert_{r_1} - \left.\frac{d^3}{dr^3}\phi_{A}^{\Ron{2}}\right\rvert_{r_1} +
	\left.2\alpha_1\frac{d}{dr}\phi_{A}^{\Ron{1}}\right\rvert_{r_1}& \nonumber \\[3pt]
	- \frac{2\alpha_1 + 4 \alpha_1(m+\alpha_1 r_1^2)}{r_1}\phi_{A}^{\Ron{1}}(r_1) &= 0, \\
\intertext{and at $r = r_2$, which is the boundary between regions \Ron{2} and \Ron{3},}
	\phi_{A}^{\Ron{3}}(r_2) - \phi_{A}^{\Ron{2}}(r_2) &= 0,  \\[6pt]
	\left.\frac{d  }{dr  }\phi_{A}^{\Ron{3}}\right\rvert_{r_2} - \left.\frac{d}{dr}\phi_{A}^{\Ron{2}}\right\rvert_{r_2} &= 0,  \\[6pt]
	\left.\frac{d^2}{dr^2}\phi_{A}^{\Ron{3}}\right\rvert_{r_2} + 2\alpha_2\phi_{A}^{\Ron{3}}(r_2) - \left.\frac{d^2}{dr^2}\phi_{A}^{\Ron{2}}\right\rvert_{r_2}  &= 0, \\[6pt]
	\left.\frac{d^3}{dr^3}\phi_{A}^{\Ron{3}}\right\rvert_{r_2} - \left.\frac{d^3}{dr^3}\phi_{A}^{\Ron{2}}\right\rvert_{r_2} +
	\left.2\alpha_2\frac{d}{dr}\phi_{A}^{\Ron{3}}\right\rvert_{r_2}&  \nonumber \\[3pt]
	- \left( \frac{2\alpha_2}{r_2} + \frac{m_2^2 - m_3^2}{r_2^3} + 2\alpha_2^2 r_2\right)\phi_{A}^{\Ron{3}}(r_2) &= 0.
\end{align}
The corresponding eigenenergies of an MQR in BLG can be obtained by finding the roots of the secular equations obtained from the boundary conditions.


\section{RESULTS AND DISCUSSION}\label{secIII}
\subsection{Magnetic quantum dot}\label{secIIIA}
\begin{figure}
    \includegraphics[width = 7.5cm]{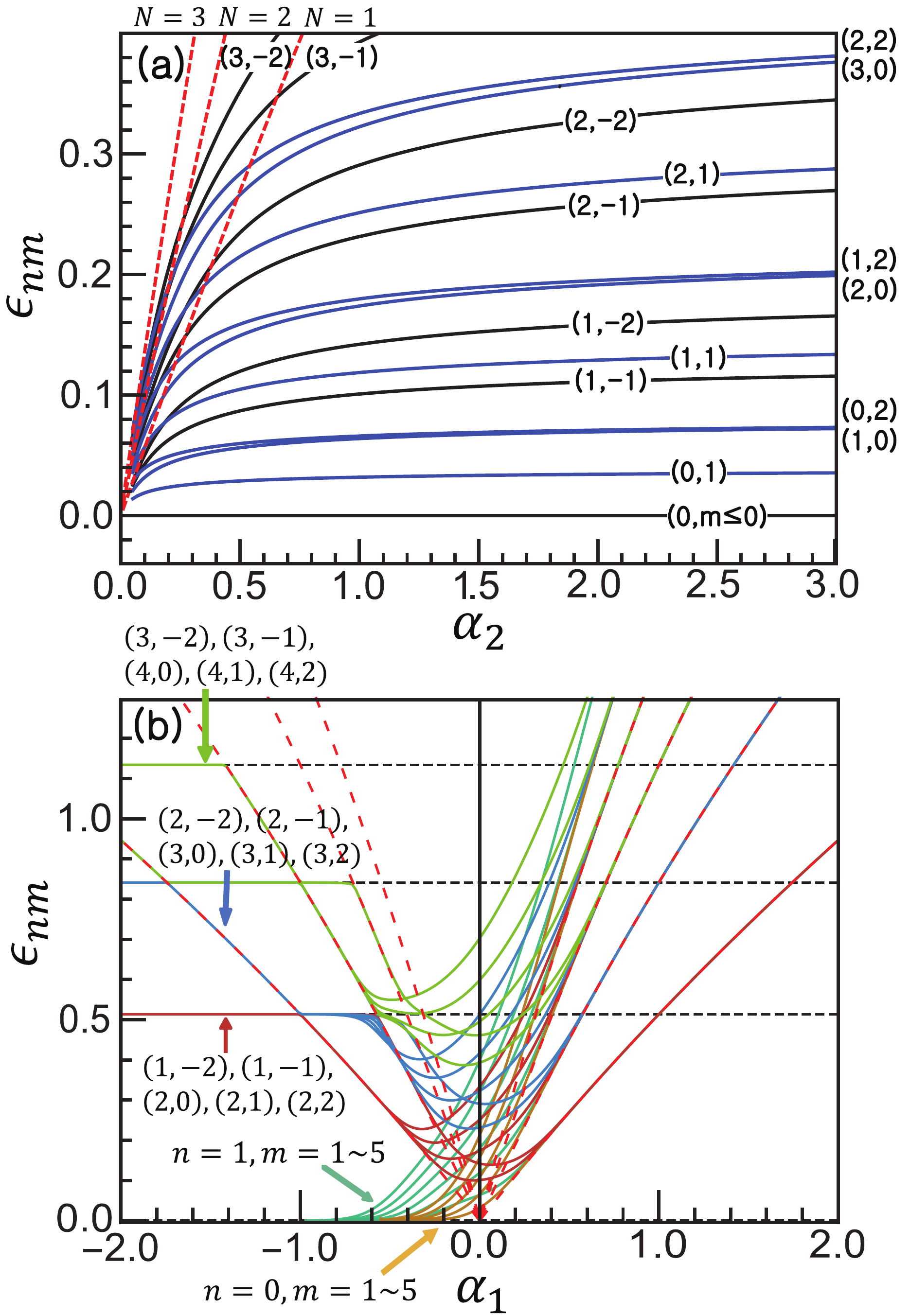}
	\caption{
		Eigenenergies $\epsilon_{nm}$ of BLG with an MQD: (a) $\epsilon_{nm}$ as  functions of
		$\alpha_2$ with $\alpha_1=0$; (b) $\epsilon_{nm}$ as functions of $\alpha_1$ with $\alpha_2=1.0$. \label{fig2}
	}
\end{figure}

In this subsection, we discuss the electronic properties of BLG with an MQD.
In Figs.~\ref{fig2}(a) and \ref{fig2}(b), the eigenenergies $\epsilon_{nm}$ are plotted as functions of $\alpha_2$ and $\alpha_1$,
respectively,  for two different MQD cases. The radius $r_0$ of the MQD is $5$ in dimensionless units.  In the quantum state $(n,m)$,  $n$ is the number of nodes in the
radial wavefunction and $m$ is the angular momentum quantum number.
First of all, to allow comparisons with the typical MQDs considered in previous work based on a 2DEG model, we consider the case in which
the inside magnetic field $\alpha_1$ of MQD is zero and the outside magnetic field $\alpha_2$ varies. The results are shown in Fig.~\ref{fig2}(a).
There is a noticeable difference in that we have zero-energy states for the MQD in BLG. These zero-energy states are also present for an MQD in monolayer graphene.
The red dashed lines in the figure represent the dimensionless Landau levels (LLs) for BLG~\cite{PhysRevB_76_115419_Pereira_2007}, i.e.,
\[
\sqrt{2\alpha}\, \sqrt{\tfrac{t^2 / \alpha + 4  N + 2}{4} - \sqrt{\left(\tfrac{t^2/\alpha + 4 N +2}{4}\right)^2 - N(N+1)}} \,,
\]
where $N$ is the Landau index corresponding to $n+(m +|m|)/2$ in our calculation.
For a magnetic field that is weak compared with  $t$, the LLs are approximately proportional to $B$.
%
Except for these features, the energy spectra of BLG with a typical MQD
show the general characteristics of an MQD in the 2DEG case~\cite{PhysRevLett_98_066802_Martino,
PhysReports_394_1_Lee, PhysRevLett_80_1501_Sim}.
For a small outer magnetic field $\alpha_2$, the wavefunctions are mostly distributed
outside the MQD, as a consequence of which the energies are close to the LLs of BLG.
On the other hand, as  $\alpha_2$ increases, the cyclotron radius of the
electrons becomes smaller than the radius of the dot, and the energies start to deviate from the LLs.
Because the Lorentz force leads to strong localization of the electrons within MQD, the energy evolution is very similar to that of an MQD in the 2DEG case. These energy changes are also the same as for a conventional circular dot that is electrostatically confined by hard
walls without magnetic fields.

In Fig.~\ref{fig2}(b),  the inside magnetic field $\alpha_1$ varies, while the outside magnetic field is fixed at $\alpha_2=1$.
The LLs  inside the MQD (red dashed lines) are fan-shaped because $\alpha_1$ increases, and the LLs  outside the MQD (black dashed lines) are flat because $\alpha_2$ remains constant.
When $\alpha_1$ increases in the positive direction, the wavefunctions are localized within the MQD. Thus, the eigenenergies approach  the LLs inside the MQD.
However, when $\alpha_1$ becomes negative, the evolution of the eigenenergy exhibits two peculiar characteristics: (i) the eigenenergies converge to the LLs outside the dot in a
stepwise manner through energy anticrossing; (ii)
the eigenenergies of some quantum states approach  zero energy. We now consider these characteristics in greater detail.
%
\begin{figure}
    \includegraphics[width = 8.9cm]{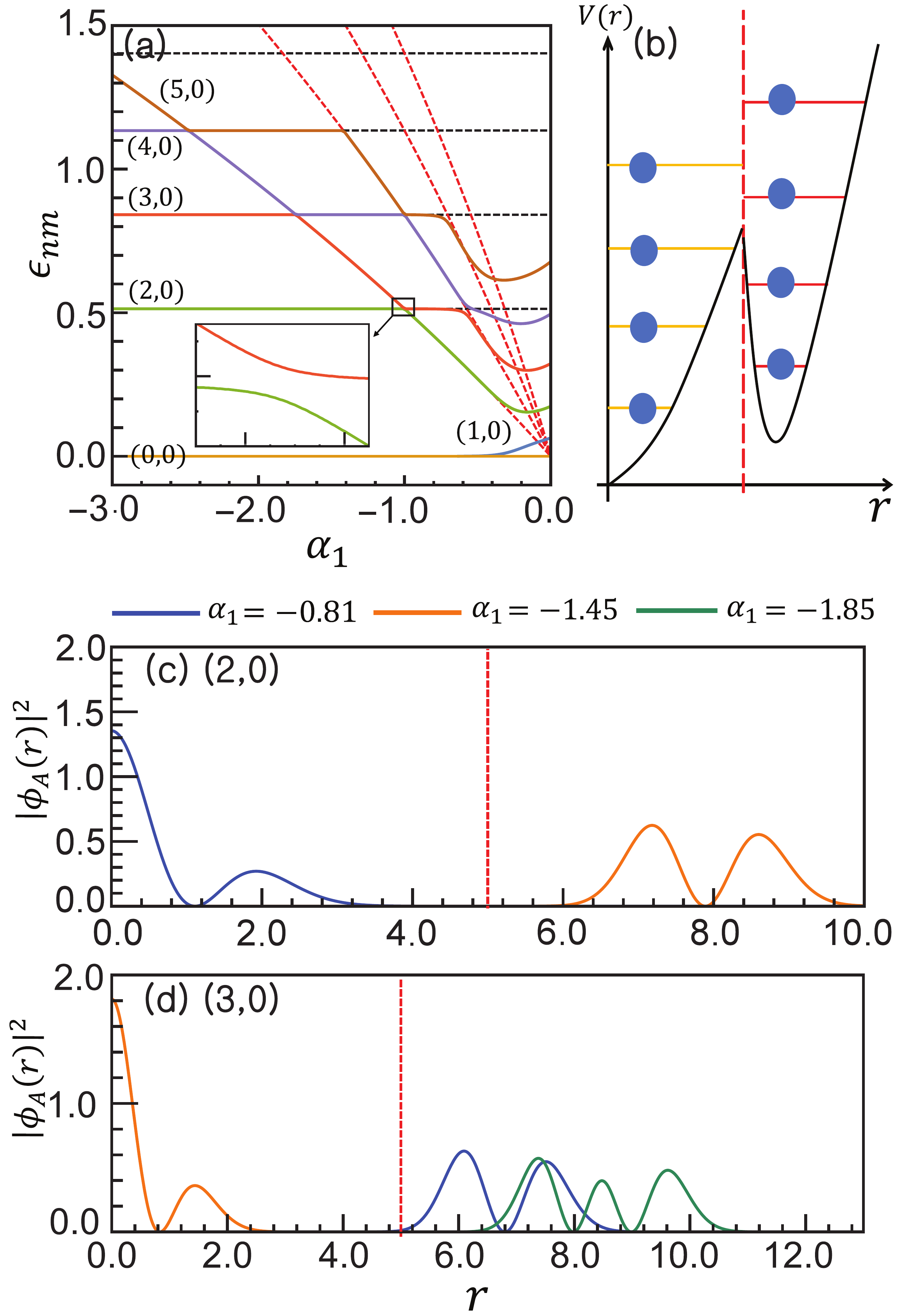}
	\caption{Characteristics of BLG with an MQD with negative inner magnetic field $\alpha_1$. (a)
	Eigenenergies $\epsilon_{nm}$ as functions of negative $\alpha_1$ for $m=0$. The inset shows energy anticrossing between two LLs.
	(b) Schematic  of effective potential for $m=0$. (c) and (d) probability densities $|\phi_A(r)|^2$ for the $(2,0)$  and
	 $(3,0)$ states, respectively. The red dashed lines in (b)--(d) indicate the radius $r_0$. All the wavefunctions are normalized.\label{fig3}}
\end{figure}

To explain the stepped shape of the energy evolution for a negative inner magnetic field $\alpha_1$, in Fig.~\ref{fig3} we plot $\epsilon_{nm}$ as
functions of negative $\alpha_1$, the effective potential of MQD for $m=0$, and the probability density of a given quantum state $(n,m)$.

In Fig.~\ref{fig3}(a), the red dashed lines and black dashed lines indicate the LLs  inside  and outside the MQD, respectively. The $(0,0)$ state has
zero energy regardless of $|\alpha_1|$, and the $(1,0)$ state tends to  zero energy as $|\alpha_1|$ increases.
The eigenenergies $(n\geq2,0)$ are determined by one of  the LLs  inside or outside the MQD, depending on the value of $\alpha_1$.
For small $\alpha_1$ near zero, the wavefunctions are  located mainly at the boundary of the MQD, and
the eigenenergies deviate from  the LLs both inside and outside the MQD. However, as $|\alpha_1|$ starts to increase, the wavefunctions
 move to the inside of the MQD and are affected by the inside magnetic field $\alpha_1$, and the energies start to follow the LLs  inside the MQD.
When an LL  inside the MQD
becomes equal to an LL  outside the MQD,
a state following an LL inside the MQD transitions to an LL outside the MQD, and vice versa.
At such a time, energy anticrossing occurs, as shown in the inset in Fig.~\ref{fig3}(a).
These anticrossing phenomena are also found in  electrostatically confined quantum dots and rings in monolayer graphene and BLG~\cite{NanoLett_9_4088_Zarenia, Appl_Phys_Lett_97_243106,PhysRevB_102_075429}.

Figure~\ref{fig3}(b) is a qualitative representation of the effective potential for the $m=0$ state of BLG with an MQD.
We shall  explain these anticrossing phenomena in terms of the effective potential.
If whole wavefunctions can be localized either inside or outside the MQD, we can simply assume that the system energies are  determined only by the LLs in the respective region.
As the magnetic field $\alpha_1$ increases, the  energy-level spacing  inside the MQD becomes wider, and the energy levels start to become equal to the LLs  outside the MQD.
When the first LL  inside the MQD becomes higher than the first LL  outside the MQD, to maintain the energy level sequence for a fixed $m$, the
state with the first LL  inside the MQD moves to the first LL  outside the MQD.
Therefore, energy anticrossing occurs whenever LLs in the two different regions become the same.
As the magnetic field increases, the wavefunction moves back and forth between the outside and inside of the MQD until the  first inside
LL becomes larger than the other fixed outside LLs, and, as a result, the wavefunctions of all states are located outside the MQD.

The $(2,0)$ state experiences  energy anticrossing once and the $(3,0)$ state three times in the region $0<|\alpha_1|<3$, as shown in Fig.~\ref{fig3}(a).
To explain the alternating movement of wavefunctions as the magnetic field inside the MQD varies, we show  the  probabilities densities
of the wavefunctions  for the $(2,0)$ and $(3,0)$ states in Fig.s~\ref{fig3}(c) and \ref{fig3}(d), respectively. For the $(2,0)$ state, the wavefunction moves
from inside the MQD at $\alpha_1=-0.81$ to outside  it at $\alpha_1=-1.45$ [Fig.~\ref{fig3}(c)]. For the $(3,0)$ state, the wavefunction moves  from
outside the MQD at $\alpha_1=-0.81$ to inside  it at $\alpha_1=-1.45$, and then moves out again at $\alpha_1=-1.85$ [Fig.~\ref{fig3}(d)].
Before and after  energy anticrossing occurs, the  probability densities are located alternately inside and outside the MQD.
\begin{figure}
\centering
    \includegraphics[width = 8.5cm]{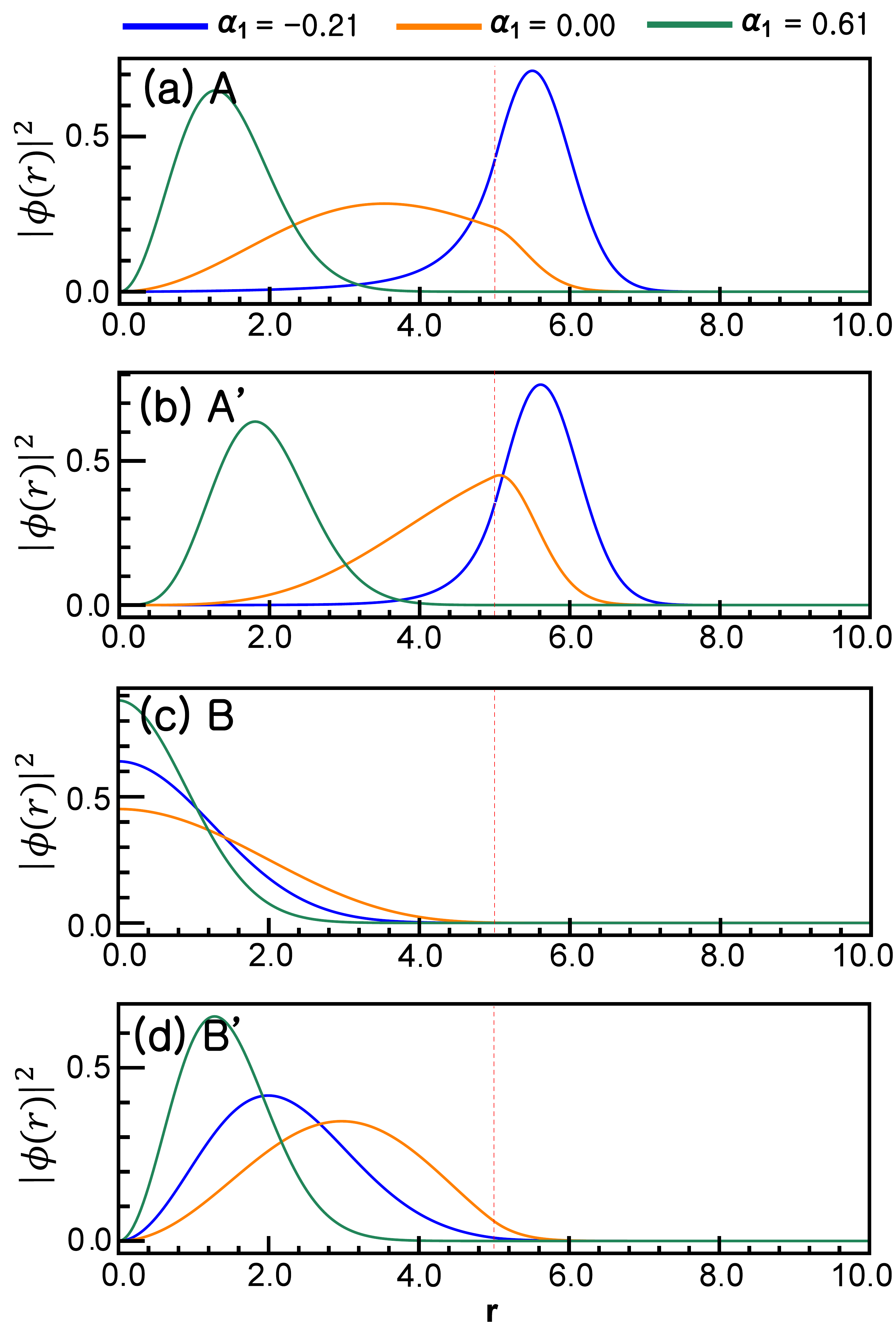}
	\caption{Probability density $|\psi(r)|^2$ for subatoms $A$, $B$, $A'$, and $B'$ in the  $(0,1)$ state for different magnetic
	fields $\alpha_1 = -0.21$,	$0.00$, and $+0.61$. To allow comparison, all the wavefunctions are normalized. The red dotted lines indicate the
	radius $r_0$.\label{fig4}}
\end{figure}

It is known that in monolayer graphene and BLG in a uniform magnetic field, wavefunctions of states at the zero Landau energy are completely localized in  a
particular subatom~\cite{PhysRevB_78_033403_Nakamura}.
For the opposite magnetic field direction, the wavefunctions are all localized in the other subatom.
In the present study, we consider a nonuniform magnetic field, and  the magnetic field inside the MQD is opposite to that outside it. We would
like to check if  wavefunction separation occurs in this case as well.
In Fig.~\ref{fig4}, we plot the probability density of different subatoms for $\alpha_1=-0.25$, $0.00$, and $+0.61$.
%
For $\alpha_1=+0.61$ (green lines), the $(0,1)$ state has nonzero eigenenergy, and the probability densities of all subatoms
are located  within the MQD in Figs.~\ref{fig4}(a)--\ref{fig4}(d). There is no wavefunction separation.
For $\alpha_1=0.00$ (orange lines), in Figs.~\ref{fig4}(a) and \ref{fig4}(b), the probability densities
of the $A$ and $A'$ subatoms are distributed over the MQD boundary, while in Figs.~\ref{fig4}(c)  and \ref{fig4}(d), the wavefunctions of the $B$ and $B'$ subatoms
are localized inside the MQD. There is still no clear wavefunction separation.
Now, for $\alpha_1=-0.21$ (blue lines),  this state approaches zero energy, and   the wavefunctions of the $A$ and $A'$ subatoms are mainly
located outside the MQD, as shown in  Figs.~\ref{fig4}(a) and \ref{fig4}(b), while the wavefunctions of the $B$ and $B'$ subatoms are more strongly localized
within the MQD than those for $\alpha_1 = 0.00$, as shown in Figs.~\ref{fig4}(c) and \ref{fig4}(d). As a result, $|\phi_A|^2$ and $|\phi_{A'}|^2$ are
distributed in a ring shape along the MQD boundary, while $|\phi_{B}|^2$ and $|\phi_{B'}|$ are distributed inside the MQD.
The $(n,m \geq 0)$ states, where $n=0,1$, tend toward zero energy, because the wavefunctions of each subatom start to be separated when $\alpha_1$ increases in the negative direction.
From these results, we can infer that  wavefunction separation also occurs  for zero-eigenenergy states in BLG,  even with a nonuniform magnetic field distribution.

\subsection{Magnetic quantum ring}\label{secIIIB}
\begin{figure}
    \includegraphics[width = 8.2cm]{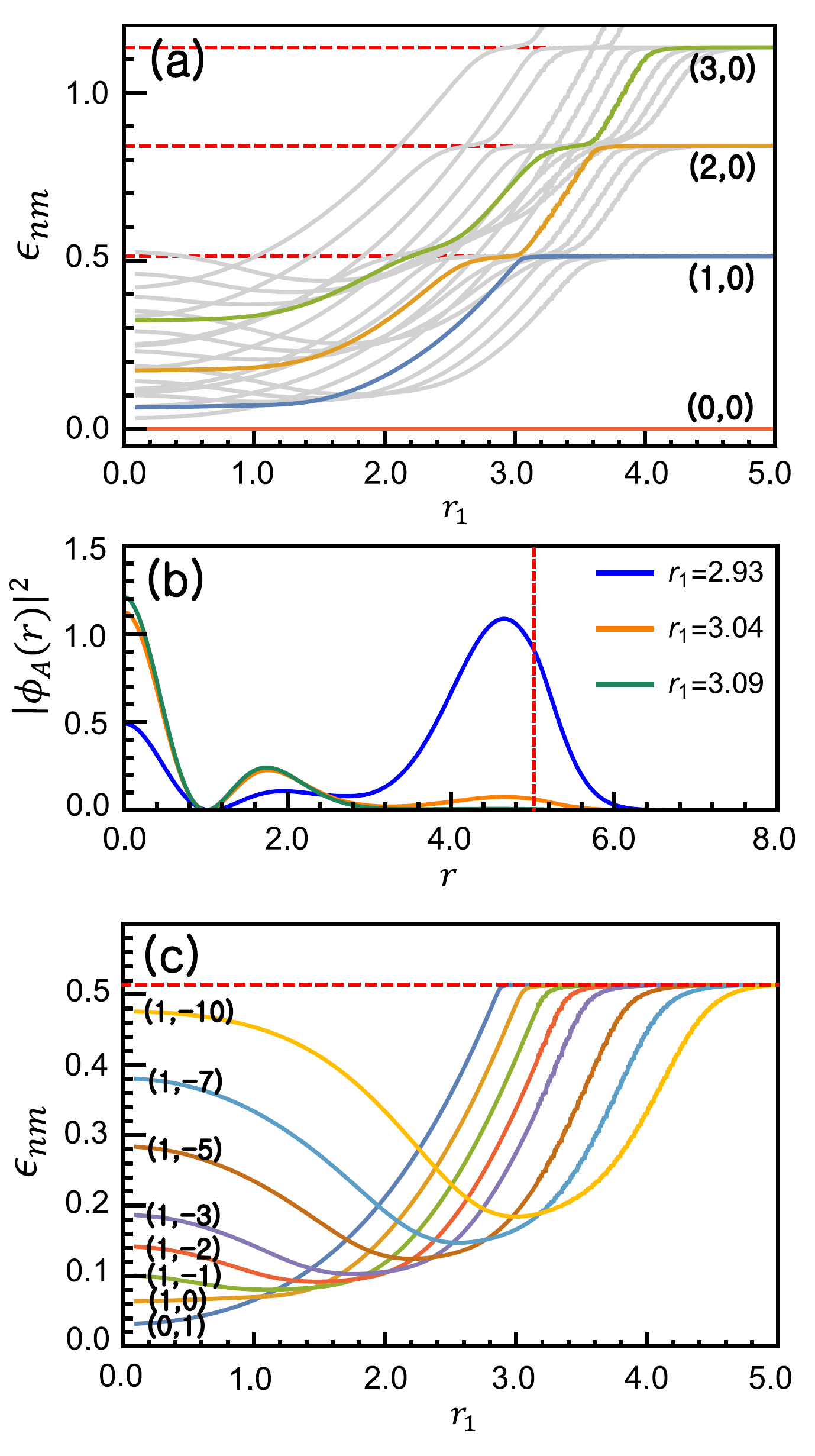}
	\caption{(a) Eigenenergies $\epsilon_{nm}$ as  functions of $r_1$ for an MQR in BLG. Colored lines are for the state with $m=0$, and gray
	lines represent other states $(n,m)$, where $n\in\AQ{[-2,2]}$ and $m\in\AQ{[-3,3]}$. Red dashed lines show the LLs of BLG.
	(b) Probability densities $|\psi(r)|^2$ of the $(1,0)$ state for different values of $r_1$ near energy anticrossing. The red dashed line indicates
	the outer radius $r_2$ of the ring. (c) $\epsilon_{nm}$ as a function
	of $r_1$ in the region of $\epsilon_{nm} < 0.5$ for the $(1,m)$ states. \label{fig5} }
\end{figure}

In this subsection, we discuss the electronic properties of BLG with an MQR.
For a fixed ring area with parallel magnetic fields ($\alpha_1=\alpha_2$ ), the results for the eigenenergy spectra show
nothing unusual compared with the cases of 2DEG and monolayer graphene~\cite{PhysReports_394_1_Lee, PhysRevB_60_8767, Phys_Lett_A_383_125865, Park2020}. Here, we investigate BLG with an MQR from a different perspective.

First, we investigate the eigenenergies by varying the inner radius $r_1$ of the MQR with parallel magnetic fields ($\alpha_1=\alpha_2$ ).
We fix the outer radius of the MQR  as $r_2=5$. The results are presented in Fig.~\ref{fig5}.
In Fig.~\ref{fig5}(a), the eigenenergies $\epsilon_{nm}$ are shown as functions of $r_1$.
The eigenenergy of the $(1,0)$ state (blue line) increases until it converges to  the LL ($n=1$) inside the ring as $r_1$ increases. The
eigenenergy of the $(2,0)$ state (orange line) exhibits a similar tendency, and it converge to the LL ($n=2$) inside the ring through energy
anticrossing near the LL ($n=1$), as shown in Fig.~\ref{fig5}(a)
 The average kinetic energy of an electron in a magnetic field is higher than in the absence of the
  field. Increasing the inner radius $r_1$ causes a decrease in the missing magnetic flux in the ring area, and so the eigenenergies of the BLG increase.


To explain the wavefunction movement near  anticrossing, in Fig.~\ref{fig5}(b), we plot the probability density distribution of the $(1,0)$ state as a function of $r$ for different  values of the inner ring radius $r_1$
For $r_1=2.93$ (blue lines) before anticrossing, the wavefunction is distributed widely over the MQR.
For $r_1=3.09$ (green lines) after anticrossing, the wavefunction is localized inside  the inner circle of the MQR.
The energy converges to the LL ($n=1$) inside the ring when  $r_1> 3.04$.
This anticrossing can be explained using the same arguments as in the MQD case.
At  anticrossing, the size of the inner magnetic field region is large enough to encompass the
wavefunctions of all subatoms for a lower energy. The wavefunctions within the ring move to the inner magnetic field region ($r_1<r$),
while a part of the wavefunction for  higher energy moves to the ring region $r_1<r<r_2$.
Before and after  energy anticrossing occurs, the wavefunctions are alternately located in regions with $B=0$ and $B\neq 0$ (inside the inner
circle or outside the outer circle) of the MQR.

To  study the angular momentum transition as the inner radius of the MQR varies, in Fig.~\ref{fig5}(c), we plot $\epsilon_{nm}$ as  functions
	of $r_1$ in the region  $\epsilon_{nm} < 0.5$ for the $(1,m)$ states.
For an MQR with a finite width and for a conventional quantum ring confined by an
electrostatic potential, an angular momentum transition is observed with increasing magnetic field~\cite{NanoLett_7_946_Pereira2007, PhysReports_394_1_Lee, PhysRevB_60_8767}. When the size of the ring is fixed, this
transition is due mainly  to the missing flux quanta in the ring area. However, in the present study,  an angular
momentum transition is observed with increasing  inner ring radius $r_1$ for a fixed magnetic field.
As the inner ring radius increases, the number of missing flux quanta in the ring area ($r_1 < r < r_2$) decreases, but the number of flux quanta within the inner ring
radius increases. The small value of $m$ leads to rapid recovery of the LL as the inner circle radius $r_1$ increases. From the existence of this phenomenon, we can
infer that the angular momentum transition depends more on the inner circle flux quanta than on the missing flux quanta in the ring area. The argument
based on missing flux quanta is only valid when the ring area remains constant.

\begin{figure}
    \includegraphics[width = 8.0cm]{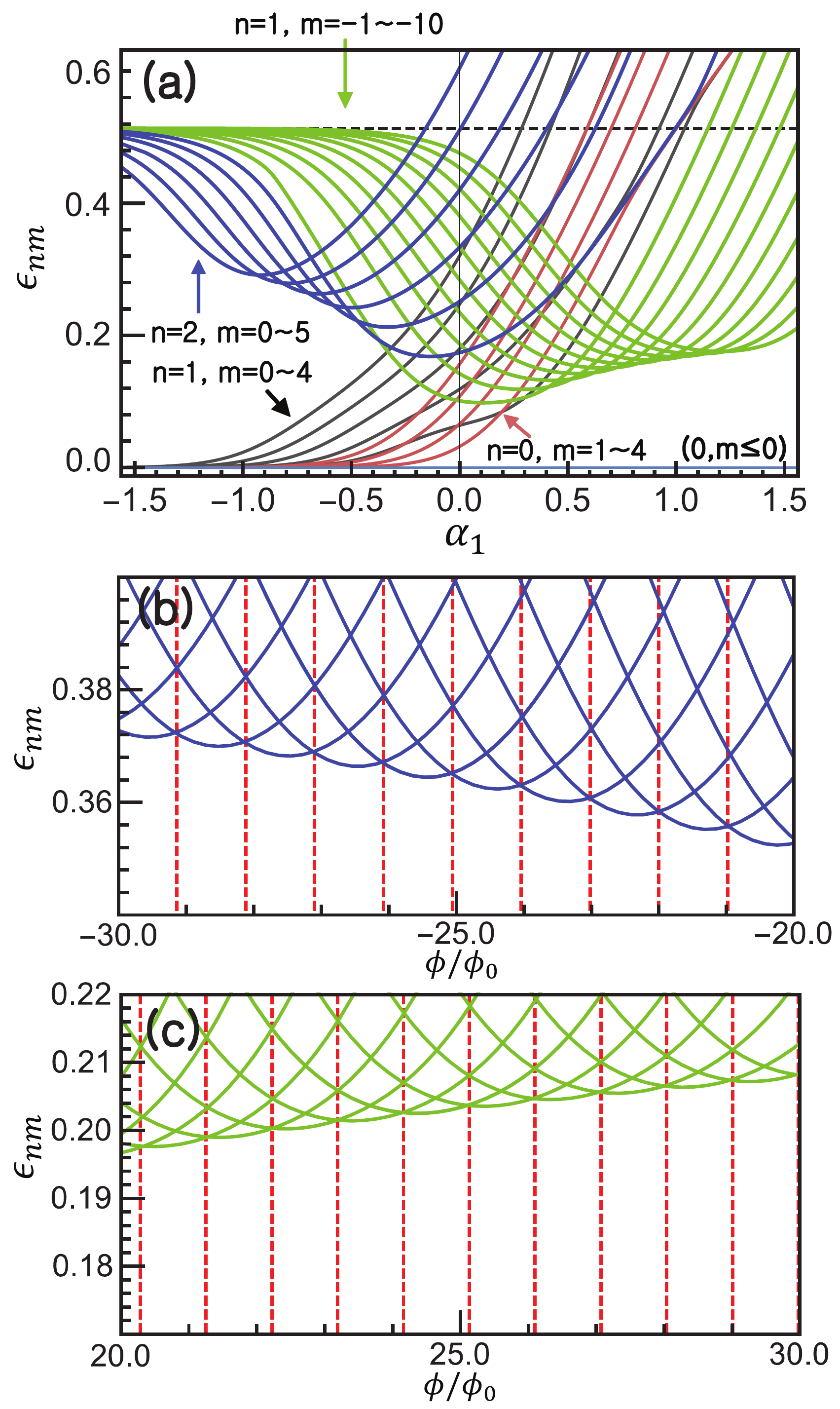}
	\caption{Energy spectra of BLG with an MQR. The inner radius $r_1$ and outer radius $r_2$ of the MQR are
	$3$ and $5$, respectively. (a) Eigenenergies $\epsilon_{nm}$ as functions of $\alpha_1$. (b) Eigenenergies  $\epsilon_{nm}$ of the states $(2,m)$, with
	$m\in [16,25]$ as functions of  magnetic flux. (c) Eigenenergies  $\epsilon_{nm}$ of the states $(1,m)$, with $m\in [-20, -30]$, as functions of magnetic flux. \label{fig6}}
\end{figure}
In Fig.~\ref{fig6}, we show the eigenenergies $\epsilon_{nm}$ of BLG with an MQR as functions of $\alpha_1$ for fixed $r_1=3$, $r_2=5$, and $\alpha_2=1$.
Here, we change the magnetic field $\alpha_1$ of the inner circle. As we vary  $\alpha_1$  from $-1.5$ to $+1.5$, we can see the energy spectra
of the MQR in parallel magnetic fields as well as in  antiparallel fields.
The energies of the $(0,m\geq0)$ and $(1,m\geq0)$ states of the
MQR approach  zero, similarly to what happens for the MQD, although the states of the MQR reach zero energy at a larger
$|\alpha_1|$ value than those of the MQD. Geometrically, the
difference between an MQD and an MQR is the presence in the latter of the zero-magnetic-field region $r_1<r<r_2$. This difference is responsible for
the very different behavior of $\epsilon_{nm}$.

For conventional quantum rings (CQRs) formed by an electrostatic potential or by an etched boundary and based on a 2DEG and monolayer graphene, the Aharonov--Bohm (AB) effect has
been studied both theoretically and experimentally~\cite{Nature_413_822_Fuhrer_2001,J_Appl_Phys_114_214314_Farghadan, PhysRevB_89_075418, J_Appl_Phys_121_024302, PhysRevB_77_085413, PhysRevB_76_235404}.
On the other hand, little attention has been paid to the AB effect for an MQR in a 2DEG.
In a CQR, the energy of an electron bound in the ring oscillates as a function of the magnetic flux $\phi$ inside the ring.
We find that the energy dispersion of BLG with an MQR also exhibits this effect for both large positive and large negative values of $|\alpha_1|$.
To investigate the AB effect for an MQR, we calculate the period of the magnetic flux $\Delta\phi $ of the MQR numerically, and  we find that  this period
is close to $\phi_0$: for $\alpha_1<0$, it is $1.01\phi_0$, and for $\alpha_1 > 0$, it is $0.97\phi_0$. For small $\alpha_1$, the wavefunctions are
not well localized in the ring region. Thus, the AB effect is not clearly seen in the range between $\alpha_1=-0.5$ and $\alpha_1=0.5$.
However, as $|\alpha_1|$ increases, the quantum
states start to be strongly localized in the ring region, and exhibit an obvious AB effect. Unlike CQRs, whose
energy has symmetry with regard to the direction of the magnetic flux inside the ring~\cite{Nature_413_822_Fuhrer_2001}, the energy in our case does not possess such
symmetry, because our MQR is formed by a nonuniform magnetic field.
We believe that this AB effect should also be seen in MQRs based on a 2DEG or monolayer graphene.

\section{CONCLUSION}\label{secIV}

In this paper, we have analytically studied the electronic properties of  bilayer graphene (BLG) with a magnetic quantum dot (MQD) and a magnetic quantum ring (MQR). We have calculated eigenenergies and wavefunctions based on a continuum model of the BLG using the Dirac equation. In contrast to previous studies, we have considered  as significant variables the  nonzero magnetic field inside the quantum dot, the size of the  dot, the radius of the inner ring, and the directions of the magnetic fields inside the inner ring and outside the outer ring.

For the MQD, as the inside magnetic field increases in the positive direction, the corresponding eigenenergies of the BLG approach the Landau levels (LLs) inside the dot, and its
states are localized  inside the  dot. However, when the magnetic field inside the  dot has the opposite direction to that
outside the  dot, we see two characteristic features. The first  is a stepwise evolution of states through energy anticrossing.
The second  is that there are quantum states approaching zero energy. In these zero-energy  states,
the wavefunctions at each subatom are separated to lie either inside or outside the MQD.
We explain these characteristics by using an effective potential and a separation of the wavefunction probability density.

For the MQR, an angular momentum transition occurs when $r_1$ is varied, and this transition depends more on the magnetic flux inside the inner circle of the ring than on the missing flux quanta in the ring area.
For a negative magnetic flux $\alpha_1$, the MQR behaves similarly to the MQD, with some quantum states approaching zero energy, although the critical value of $\alpha_1$ associated with zero energy is larger than
in the case of an MQD.
We have investigated the Aharonov--Bohm effect on energy spectra of the MQR.  For a large magnetic flux inside the ring, the period of the Aharonov--Bohm effect in
the ring is close to the value of the magnetic flux quantum $\phi_0$.

We believe that our results here could be helpful as a basis for further  study of quantum transport in BLG with these magnetic quantum structures.

\begin{acknowledgments}
This work was supported by Basic Science Research Program through National Research Foundation of Korea(NRF) funded by the Ministry of Science and ICT (Grant No. NRF-2019R1A2C1088327) and the Ministry of Education (Grant No. NRF-2018R1D1A1B07046338).
\end{acknowledgments}

\bibliography{kim_ref_MQR.bib}
\bibliographystyle{apsrev4-2}
\end{document}